\documentclass[sigconf]{acmart}
\usepackage{graphicx} 
\usepackage{colortbl} 
\usepackage[table]{xcolor} 
\usepackage{booktabs} 
\usepackage[capitalise]{cleveref}
\usepackage[most]{tcolorbox}
\usepackage{enumitem} 
\usepackage{tabularx}
\usepackage{balance}
\usepackage{hyperref}

\tcbset{
  interviewquote/.style={
    enhanced,
    breakable,
    colback=orange!5,
    colframe=orange!20,
    left=0.2em,
    right=0.2em,
    top=0.1em,
    bottom=0.1em,
    boxrule=0pt,
    borderline west={1pt}{0pt}{orange!60},
    sharp corners,
  }
}

\newcommand{\interviewquote}[2]{%
  \begin{tcolorbox}[interviewquote]
    \emph{``#1''} #2
  \end{tcolorbox}
}

\usepackage{fbox}
\newcommand{\summary}[2]{
        \vspace{2mm}
        \noindent
        \fbox{%
            \parbox{.97\linewidth}{%
                    \textbf{#1 Summary.}
                #2
            }%
        }%
}%

\AtBeginDocument{%
  }

\copyrightyear{2026}
\acmYear{2026}
\setcopyright{cc}
\setcctype{by-nc-nd}
\acmConference[FSE Companion '26]{34th ACM Joint European Software Engineering Conference and Symposium on the Foundations of Software Engineering}{July 05--09, 2026}{Montreal, QC, Canada}
\acmBooktitle{34th ACM Joint European Software Engineering Conference and Symposium on the Foundations of Software Engineering (FSE Companion '26), July 05--09, 2026, Montreal, QC, Canada}
\acmDOI{10.1145/3803437.3805791}
\acmISBN{979-8-4007-2636-1/2026/07}

\begin{document}

\title{Domain Diversity, Motivation, Inclusion, and Feedback in Software Modelling Education}

\author{Isabella Graßl}
\email{isabella.grassl@tu-darmstadt.de}
\affiliation{%
  \institution{TU Darmstadt}
  \city{Darmstadt}
  \country{Germany}
}

\author{Christopher Lazik}
\email{lazikchr@hu-berlin.de}
\affiliation{%
  \institution{HU Berlin}
  \city{Berlin}
  \country{Germany}
}

\author{Shalini Chakraborty}
\email{s.chakraborty@uni-bayreuth.de}
\affiliation{%
  \institution{Universität Bayreuth}
  \city{Bayreuth}
  \country{Germany}
}

\author{Grischa Liebel}
\email{grischal@ru.is}
\affiliation{%
  \institution{Reykjavik University}
  \city{Reykjavík}
  \country{Iceland}
}

\author{Miguel Goulão}
\email{mgoul@fct.unl.pt}
\affiliation{%
  \institution{NOVA-LINCS, NOVA School of Science and Technology}
  \city{Lisbon}
  \country{Portugal}
}

\renewcommand{\shortauthors}{Graßl, Lazik, Chakraborty, Liebel and Goulão}

\begin{abstract}
Student engagement is critical for effective learning in software modelling, yet fostering motivation and inclusivity remains a challenge. While existing research has focused on modelling tools, notations, and assessment, little attention has been given to how the choice of problem domains and the diversity, relatability, and cultural perspectives they bring shape students’ learning experiences.
This study explores how problem domains and teaching methods influence motivation, engagement, inclusiveness, and feedback in modelling education. To investigate these dimensions, we conducted parallel surveys with 90 students and 22 educators. 
Our findings reveal disconnects between educator assumptions and student preferences: Students show greatest motivation for socially relevant domains and prefer choice in selection, while educators overestimate interest in study-related domains. The study identifies how minor design choices can exclude students. Students perceive feedback as meaningful when visibly acted upon. These findings suggest inclusive domain selection is central to student motivation; thus, we recommend student-centred domain selection.
\end{abstract}



\maketitle

\section{Introduction}
Software modelling is a core skill in software engineering (SE) education, enabling students to abstract, conceptualise, design, and manage complex systems \cite{cowling2005role,hutchinson2011empirical}. 
Despite its importance, software modelling has often been under-emphasised in curricula \cite{cowling2003modelling}. 
Students may find modelling conceptually demanding \cite{chakraborty2023we}, struggle with tools \cite{agner2019student,liebel2016impact} and perceive it is as detached from practical work \cite{badreldin2015effects}. Educators, meanwhile, face limited class time and resources, and modelling is often taught in a top-down and formal way \cite{whittle2014studio}, making student engagement a challenge.

One promising way to motivate students in modelling education is the choice of example domains in assignments and lectures, as Paige et al.~\cite{paige2014bad} observed: \emph{``students both need and benefit from examples when learning how to model. Examples reinforce the conceptual principles and engineering practice of modelling”}. 
The choice of domains is not merely illustrative since relatable and interesting domains might help students connect abstract concepts to what they value. In contrast, traditional examples such as games or library management systems may not resonate with all students and can reduce engagement, contribute to exclusion, and harm learning outcomes~\cite{scott2015}. 
However, problem domains are only one part of the learning experience. The way modelling is taught shapes how students engage with it. Teaching methods such as collaborative~\cite{maruyama2018support} or gamified~\cite{bucchiarone2023gamifying,garaccione2024gamification} learning can make modelling more meaningful.

Recent work highlights the importance of human factors in software modelling \cite{liebel2024human} and inclusion in modelling education \cite{bork2025inclusive}, aiming to create classrooms where all students feel welcomed and represented. 
However, while there is substantial research on software modelling education on tools \cite{akayama2013tool,liebel2016impact,agner2019student}, student learning \cite{bach2022analysis,maier2024uml,ardimento2024enhancing}, assessment practices \cite{hamann2024towards,chen2024embedding}, and classroom practices \cite{fuksa2021mini,khandoker2023interdisciplinary}, there is, to our knowledge, no work that examines how the choice of problem domains influences students. 
In this study, we investigate how problem domains and teaching methods influence motivation, engagement, inclusiveness, and feedback in modelling education. We address the following research questions (RQs):\\

\noindent\textbf{RQ1:} \emph{How do different modelling domains affect student motivation and engagement?}

\smallskip
\noindent\textbf{RQ2:} \emph{How do preferences for teaching methods and assignment design choices affect motivation in modelling assignments?} 

\smallskip
\noindent\textbf{RQ3:} \emph{What are the perceptions of inclusiveness in modelling assignments?} 

\smallskip
\noindent\textbf{RQ4:} \emph{What are the perceptions of feedback in modelling assignments?} 
\smallskip

To answer the RQs, we conducted a survey with 22 educators and 90 students with modelling experience. 
Our findings show disagreement among students about which problem domains are motivating. While educators hold strong assumptions about which domains motivate students, students show much more diverse preferences driven primarily by personal interests and social relevance.
This is the first study to compare educators' and students' perspectives on motivation, engagement, inclusiveness, and feedback in modelling assignments, providing insights that can help educators design more engaging and inclusive modelling courses.

\section{Background and Related Work}
\label{sec:background}
We first examine challenges in teaching modelling, then explore what motivates students in this context, and finally consider how inclusion in teaching can address  motivational barriers.

\subsection{Software Modelling Education}
While modelling is part of many SE curricula, pedagogical emphasis often remains superficial, focused on syntax rather than conceptual modelling and model-driven thinking. \citeauthor{whittle2013state} \cite{whittle2013state} demonstrate the importance of hands-on modelling exercises, while critiquing the predominant focus on syntactic correction over real-world problem-solving. 
Educators in the MODELS community have repeatedly emphasised the lack of comprehensive teaching resources and well-established pedagogical approaches for modelling and model-driven engineering (MDE) education \cite{clarke2018teaching}. \citeauthor{gogolla2018teaching} \cite{gogolla2018teaching} conceptualise modelling education as an ecosystem, identifying the factors that constitute this ecosystem and describing their relationships. They highlight elements such as teaching domains and instructional styles as particularly influential in shaping effective modelling education.
Modelling tools commonly used in education introduce {\em accidental complexity}: setup, configuration issues, and poor usability for classroom settings, distract students from understanding core modelling concepts \cite{agner2019student,liebel2016impact}. Recent guidelines emphasise the need for teaching-specific modelling tools with easy setup, low friction, and meaningful feedback \cite{kienzle2024requirements}. 
The field's reliance on the Unified Modelling Language UML as the dominant modelling language is well-established in \citeauthor{seidl2015uml}'s \cite{seidl2015uml} instructional methodology. However, \citeauthor{petre2013uml} \cite{petre2013uml} provide empirical evidence that this academic focus does not match industrial practice, where UML adoption is often partial or tool-driven. This industry-academia gap is further quantified by \citeauthor{ciccozzi2019execution} \cite{ciccozzi2019execution}, whose analysis of executable UML models reveals significant limitations in translating educational models to production systems.
Empirical studies document students' struggles with fundamental modelling concepts such as abstraction, multiplicity, and mapping textual requirements to diagrams, highlighting prevalent misconceptions that hinder learning unless actively addressed \cite{chren2019mistakes}.
Collectively, these challenges create substantial technological and cognitive barriers for students, which in turn diminish their motivation.
This paper suggests concrete mitigation strategies for these challenges.
\subsection{Student Motivation in Software Modelling}
Motivation is critical in engineering education and can be framed through, e.g., Self-Determination Theory \cite{trenshaw2016using}, which posits that autonomy, competence, and relatedness enhance motivation and learning. 
In modelling education, motivational barriers are often tied to tooling issues, delayed or unhelpful feedback or unclear expectations from the modelling task itself \cite{akayama2013tool,kienzle2024requirements,chakraborty2023we}. 
Prior work shows that lack of experience in the subject matter limits students’ ability to engage flexibly in modelling, and that when the domains are unfamiliar, this reduces confidence and contributes to frustration \cite{cevikbas2023advantages}.
Some educators have turned to gamification, using game elements like badges or challenges, to boost engagement in modelling courses \cite{buckley2016gamification}, though results vary depending on implementation.
However, there is limited research examining the influence of problem domains in software modelling education, and particularly a lack of diversity in the example domains used.
The results in this paper help bridge this gap.

\subsection{Diversity in Problem Domains and Inclusion}
Diversity in problem domains means exposing students to a variety of contexts and applications in modelling assignments, while inclusion ensures that these domains are relatable and do not privilege specific backgrounds or prior knowledge, contributing to culturally responsive computing~\cite{scott2015,charleston2025}.
Inclusive pedagogies such as Universal Design for Instruction (UDI), culturally relevant teaching, and equitable assignment design advocate for multiple means of representation, flexible assessment, and structural responsiveness to diverse learner needs \cite{burgstahler2009universal,ladson1995but,estefan2023inclusive}. 
Active learning methods, such as model construction through guided, incremental tasks, have shown promise in improving engagement and conceptual grasp in UML contexts \cite{unkelos2019climb}.
\citeauthor{perez2020project} \cite{petre2013uml} demonstrate that project-based learning (PBL) in UML education leads to higher scores on UML examinations compared to traditional learners, suggesting active engagement in modelling projects improves knowledge retention. PBL also enables the development of transferable skills, including teamwork and communication, suggesting it can increase student motivation without compromising performance in other course components.
This aligns with \citeauthor{szmurlo2006teaching}'s \cite{szmurlo2006teaching} project-based framework, which shows iterative modelling tasks improve conceptual understanding compared to lecture-based approaches.

Prior work highlights persistent challenges in teaching software modelling, which range from limited pedagogical resources and motivational barriers to insufficiently diverse and inclusive assignment domains, yet little is known about how these factors intersect to shape student engagement and learning. This paper offers insights into the role of these factors in teaching software modelling.

\section{Method}\label{sec:method}
We investigate how the choice of problem domains and teaching methods influence motivation, engagement, inclusiveness, and feedback in modelling education and how these 
vary across diverse student groups. 
We compare students' and educators' perceptions, 
to identify potential misconceptions or biases in teaching materials.

\subsection{Instrumentation}
\begin{table*}[t]
\centering
\caption{Survey questions for students and educators. (SC/MC = single/multiple choice, FT = free text, LI = Likert)}
\label{tab:surveymerged}
\begin{tabularx}{\linewidth}{llll}
\toprule
Var. & Question & Type & Target \\
\midrule

\multicolumn{4}{l}{\textbf{Demographics}} \\ 
DE01 & What age are you? & SC & Both \\
DE02 & What gender do you identify as? & SC & Both \\
DE04 & In which country are you currently studying/teaching? & FT & Both \\
DE06 & What degree are you currently pursuing? & SC & Students \\
DE07 & How many years of work experience in software engineering do you have? & SC & Students \\
DE08 & How many years of teaching experience in software engineering do you have? & SC & Educators \\
DE09 & Which of the following minorities do you belong to? & MC & Both \\
DE10 & Which courses/topics do you typically teach? & FT & Educators \\
DE11 & Are you an active member in the modeling research community? & SC & Educators \\
DE12 & How many years of experience in software modelling do you have? & SC & Students \\
\midrule

\multicolumn{4}{l}{\textbf{Motivation $\rightarrow$ RQ1}} \\
MO01 & How motivated are you/do you think students are to work on projects related to the following\\
& domains? & -- & Both \\
MO01-01 & Video Games & 5-pt LI & Both \\
MO01-02 & Community Platforms (e.g., sharing applications, NGO platforms, social media) & 5-pt LI & Both \\
MO01-03 & Public Information Systems (e.g., traffic routing, flight booking) & 5-pt LI & Both \\
MO01-04 & Enterprise \& Business Systems (e.g., financial reporting, human resource management) & 5-pt LI & Both \\
MO01-05 & Automation Systems (e.g., smart buildings, cyber-physical systems) & 5-pt LI & Both \\
MO01-06 & Others & 5-pt LI & Both \\
MO02 & Please briefly explain your choice in the previous question. & FT & Both \\ 

\midrule
\multicolumn{4}{l}{\textbf{Design of Assignments $\rightarrow$ RQ2}} \\
DA01 & When designing modelling assignments, how do you select the domain? & FT & Educators \\
DA02 & What domains do you typically use in assignments and/or lectures? & FT & Educators \\
DA04 & How do you consider student interest or motivation when designing modelling assignments? & FT & Educators \\
\midrule

\multicolumn{4}{l}{\textbf{Preferences for Teaching Methods $\rightarrow$ RQ2}} \\
PR02 & To what extent do you think gamification in education motivates students? & 5-pt LI & Both \\
PR03 & Please briefly explain your choice in the previous question. & FT & Both \\
PR04 & Which kind of collaboration do you prefer in modelling assignments? & SC & Both \\
PR05 & Please briefly explain your choice in the previous question. & FT & Both \\
PR06 & How do you encourage students to engage with your modelling assignments? & FT & Educators \\

\midrule
\multicolumn{4}{l}{\textbf{Inclusion and Bias $\rightarrow$ RQ3}} \\
IN02 & Have you ever felt that an assignment topic was excluding a specific sub-group of students? & FT & Students \\
IN03 & Do you think the choice of problem domains in modelling assignments affects the inclusiveness of the \\
& course? & FT & Students \\
IN04 & How do you feel your learning success is impacted by the choice of domain in modelling assignments? & FT & Students \\
IN01 & Please provide examples of problem domains you believe are interesting for a diverse group of students. & FT & Students \\

EI01 & How do you consider student diversity when selecting domains for modelling assignments? & FT & Educators \\
EI02 & Have you observed any differences in students’ preferences for different modelling domains? & FT & Educators \\
EI03 & Do you have any suggestions for making modelling education more inclusive? & FT & Educators \\
\midrule

\multicolumn{4}{l}{\textbf{Feedback $\rightarrow$ RQ4}} \\
FE03 & Do you have the feeling that your feedback can impact the quality of a course? & FT & Students \\
FE01 & How do you assess and provide feedback on students’ assignments? & FT & Educators \\
FE02 & How do you evaluate the quality of your courses? & FT & Educators \\
\bottomrule
\end{tabularx}
\end{table*}
We conducted two surveys: one for students and one for educators. Both surveys were designed to capture the current state of problem domains in software modelling assignments and how these domains influence motivation, engagement, inclusiveness, and feedback in modelling education. We relied on self-reported data to capture subjective perceptions, as lived experience and personal voice were central. Although self-reports may not fully align with objective outcomes, they are often closely linked to learning behaviours and achievements~\cite{han2022,pardo2016}.

We conducted a pilot study with three students and two educators from Germany, Iceland and Portugal, to help us evaluate the questionnaire clarity. 
The pilot participants were not part of the final study.
Based on this feedback, we made minor adjustments to the wording of some questions and added one additional demographic question. 
The final version of the questionnaire included four thematic categories based on the four RQs, each focusing on different aspects of participants' experiences with modelling assignments. These categories covered topics such as motivation, preferences and design choices of assignments, inclusion, and feedback.

Before participants completed the questionnaire, we explained the study's purpose and process. We informed them of the survey's voluntary nature, emphasising they could stop at any time and that all responses were anonymous.
We did not require ethics approval since the study involved anonymous, non-sensitive questions, in accordance with the legal framework in Germany. 

\Cref{tab:surveymerged} shows the questionnaire for both students and educators. 

\paragraph{Demographics} 
To better understand our participants, we collected basic demographic data. This included internal diversity factors such as age (\emph{DE01}), gender (\emph{DE02}), country of study or teaching (\emph{DE04}), and whether they identified with a minority group, such as ethnic minorities or LGBTQ+ (\emph{DE09}). We gathered information on the participants’ educational and teaching experiences (\emph{DE06,07,08,10,12}). Educators were asked if they are active members of the research community (\emph{DE11}).

\paragraph{Motivation} 
To explore how problem domains influence motivation, we asked participants about their experiences with different modelling domains. The six domain categories (\emph{MO01-01 to MO01-06}) were derived from an analysis of the authors' teaching materials as well as 31 papers from the EduSymp workshop at MODELS conference between 2019 and 2023 since it is a well-established venue that reflects current educational practice in modelling.
We classified the domains based on the types of models students were required to create, the types of systems being modelled, and the tools used.
For each year, one of the authors reviewed the selected papers to identify the model types, system types, and tools. All authors discussed the findings and grouped the domains into six final categories.

\paragraph{Preferences in Teaching Methods (Gamification, Collaboration)}
We aimed to identify the preferences of both students and educators regarding different teaching methods. Based on our analysis of teaching materials and the 31 EduSymp papers, we found that gamification is often suggested as a potential method (e.g., \cite{bucchiarone2021merge}). Additionally, it is closely linked to student motivation \cite{tonhao2023gamification}. Before asking participants about their opinions on the use of gamification in modelling education (\emph{PR02}), we provided a brief explanation of the concept to ensure understanding.
We also explored preferences regarding collaboration in modelling assignments, asking participants whether they preferred working in groups, individually, or had no preference (\emph{PR04}). We included this question based on our observations as teachers that group work can be a major (de-)motivator. To gain further insights, participants were encouraged to justify their choices through an open-ended question.
Additionally, we asked educators for their perspectives on what motivates students to engage with modelling assignments (\emph{PR06}).

\paragraph{Design of Assignments}
We asked educators three open-ended questions about their approach to designing assignments. We inquired about how they select problem domains (\emph{DA01}), which domains they typically use (\emph{DA02}), and whether they take students' interests into account when designing assignments (\emph{DA03}). These align with choosing socially relevant domains~\cite {layman2007note} and emphasize the benefits of allowing students to select their own topics~\cite{kazerouni2025topic}.

\paragraph{Inclusion and Bias}
Our analysis of teaching materials and the 31 EduSymp papers revealed that most approaches do not explicitly address diversity and inclusion aspects.
To address this gap, we asked students whether they feel excluded from certain topics (\emph{IN02}), whether the choice of problem domains influences the inclusiveness of the course (\emph{IN03}), and whether they believe these domains affect their learning success (\emph{IN04}).

For educators, we asked how they consider diversity when designing assignments (\emph{EI01}) and whether they have observed differences in student preferences based on diversity factors (\emph{EI02}). We also asked both students and educators to suggest topics they believe would appeal to a diverse group of students (\emph{IN01, EI03}).

\paragraph{Feedback}
Feedback is crucial for improving teaching methods and assignments. We asked students whether their feedback had an impact on the quality of the course (\emph{FE03}).
We also asked educators how they assess feedback on student assignments (\emph{FE01}) and how they incorporate this feedback into their teaching practices (\emph{FE02}).

\subsection{Data Collection and Participants}
Our target group includes university students enrolled in computer science or related fields of study. To participate, students must have completed courses that covered software modelling to some extent, such as UML or Business Process modelling. Similarly, educators who have taught courses that included software modelling were invited to participate. 
The data collection took place between February and August 2025. 

\emph{Educators.} We sent an invitation to our educator survey to 117 educators via email in early February 2025. These invitees were selected from past program committees of the Models conference and authors of the EduSymp proceedings. Of the 117 invitations sent to professionals, 22 educators agreed to participate in the study, while 5 email addresses were undeliverable.

\emph{Students.} We invited students 
from the authors' universities (i.e., \emph{Bayreuth, Darmstadt, Nova, Reykjavík}) to fill in the student survey.

Overall, our dataset includes 112 participants, where 90 responses are from students and 22 from educators.

\subsection{Data Analysis}
To address our research questions, we used both quantitative and qualitative analysis methods.

\subsubsection{Quantitative Analysis}
For the closed-ended survey items, we conducted statistical analyses separately for students and educators. Ratios were calculated to provide descriptive insights into the distribution of responses. To test for statistically significant differences, we applied Fisher’s exact test (Fisher) as implemented in \texttt{R}, using a significance level of $\alpha = 0.05$, as it is appropriate for small sample sizes and categorical data.
We tested for differences in MO01-01 to MO01-05 and in PR02 between students and educators, by student gender, and by student minority status.

\subsubsection{Qualitative Analysis}
For the open-ended responses, we adopted content analysis~\cite{krippendorff2018content} and thematic analysis~\cite{braun2022} depending on the type of open-ended question. 
To support collaboration, we used a shared Google Sheets document to store all anonymised responses. 
Three researchers independently reviewed a subset of 20\% of student and educator responses to agree on the suitability of the qualitative method. 
The remaining responses were divided among the researchers based on the research questions, who coded them individually. After completing the individual coding, we triangulated the review process by having a second researcher review the coding. 
We resolved disagreements between researchers through structured discussion in joint online meetings, where they compared interpretations, justified their coding decisions with reference to the data, and refined code definitions until consensus was reached. 
This approach follows established practices in qualitative research that emphasise the collaborative construction of meaning rather than statistical agreement measures~\cite{braun2022,oconnor2020a}.

We analysed factual or list-oriented responses (e.g., typical domains used)  using \emph{content analysis} and coded them into concrete categories. 
This includes the variables 
\emph{DA02, EI02, FE01, FE02, FE03} (\cref{tab:surveymerged}).
In contrast, we analysed responses that provided reasoning, reflections, or broader perspectives (e.g., rationales for domain selection, considerations of diversity, or strategies to encourage student engagement) using \emph{thematic analysis} to identify recurring patterns and overarching themes. This includes the following variables: \emph{MO02, DA01, DA04, PR03, PR05, PR06, EI01, EI03}. 
This mixed strategy ensured that each variable was analysed with a method appropriate to the data's richness. 
Given the domain-specific nature of the questions and the relative consistency of responses, the dataset was considered sufficient to support meaningful qualitative interpretation without the aim of statistical representativeness.

\subsection{Threats to Validity}
\emph{Construct validity.} A potential threat lies in the design of our survey instruments. While we based our questions on prior work, conducted a pilot study, and iteratively refined them, some constructs (e.g., inclusiveness) may not be fully captured through self-reported data. To mitigate this, we used both closed and open-ended questions and triangulated across students’ and educators’ perspectives.  

\emph{Internal validity.} Researcher bias in the qualitative coding process may influence responses interpretation, so three researchers independently coded a subset of responses, discussed disagreements, and agreed on a shared coding scheme before coding the full dataset.  

\emph{Researcher positionality.} The backgrounds of the research team may have shaped the framing of the study and the interpretation of data. Among the five authors, two identify as women and three as men; four are based in Europe and one in Asia. Three of the authors are active modelling educators, which provided valuable contextual knowledge but may also have introduced implicit assumptions about teaching practices. We sought to mitigate these influences by engaging in collaborative coding, reflexive discussions, and by making our analytic process transparent.  

\emph{Conclusion validity.} Our analyses may be affected by limitations in statistical power (e.g., small subgroups of participants) or the potential for Type I/II errors. 
We used appropriate statistical methods and complemented quantitative analyses with qualitative insights.

\emph{External validity.} The generalisability of our results is limited by our sample of students and educators, which may not be representative of the broader population. We mitigated this by including participants from diverse institutions and backgrounds, and by clearly reporting demographic information.

\section{Results}
\label{sec:results}
This section presents the findings for each research question. As the questionnaire was fully optional to align with ethical considerations, totals across demographic, quantitative, and qualitative results may differ, as some participants chose not to answer all questions.

\subsection{Demographics}
\subsubsection{Students} 
From students, we collected 90 valid answers (25 women, 55 men, 3 non-binary, 1 other and 6 did not answer).
Of these, 62 had a bachelor's degree and 22 a master's degree.
The majority ($n=47$) had no relevant work experience, followed by 18 students with less than 1 year of experience and 17 students with 1 to 5 years of experience. 
Similarly, 49 students declared to have no software modelling experience, 19 students had less than 1 year of modelling experience, and 15 students had 1 to 5 years of experience.
However, we recruited only students who had taken at least one university-level course containing modelling topics, ensuring some exposure to these topics.
41 students declared that they belonged to at least one minority: 18 students declared to belong to an ethnic minority, 13 declared to belong to LGBTQI+, 4 declared to be disabled, 5 to be socio-economically disadvantaged, and 11 students declared to belong to another minority.

\subsubsection{Educators} 
We recruited 22 educators, 19 of whom indicated that they are also active in modelling research.
The sample consisted of 17 men and 4 women (and 1 who did not answer \textit{DE02}).
One participant indicated belonging to a gender minority, and two indicated belonging to another minority (without a free-text description).
The educators' ages covered a broad range from 25-34 ($n=5$), 35-44 ($n=7$), 45-54 ($n=4$), to 55 or older ($n=5$). 
%
Most courses (\emph{DE10}) taught by educators in our sample relate to core SE principles such as Software Engineering, Requirements Engineering, Project Management, Software Modelling and Design (e.g., UML, BPMN, design patterns). 
Some focus on programming and implementation, like object-oriented programming, and domain-specific languages.

\subsection{RQ1: Motivation}
\begin{figure}[t]
\centering
\includegraphics[width=\linewidth]{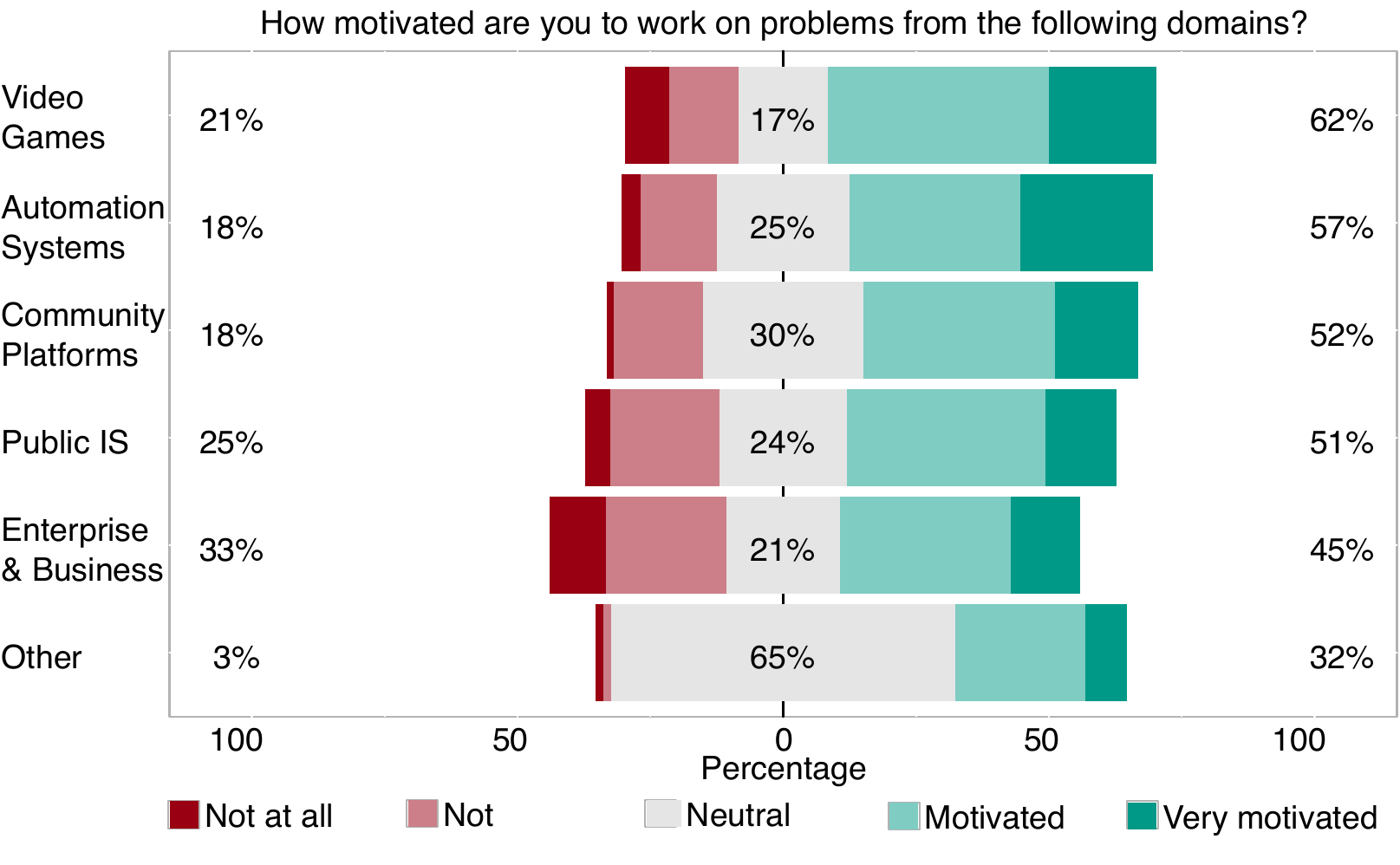}
\caption{Students' motivation ratings by domain (RQ1).}
\Description{Video games are most popular among students, with 62\% finding them motivating, followed by automation systems (57\%), community platforms (52\%), public information systems (51\%), enterprise and business (45\%), and others (32\%). However, all these domains are also perceived as not motivating by students, ranging from 18\% (Community platforms and automation systems) to 33\% (Enterprise and business systems). Only other domains are seen neutral, with only 3\% finding them not motivating.}
\label{fig:motivationStudents}
\end{figure}

\begin{figure}[t]
\centering
\includegraphics[width=\linewidth]{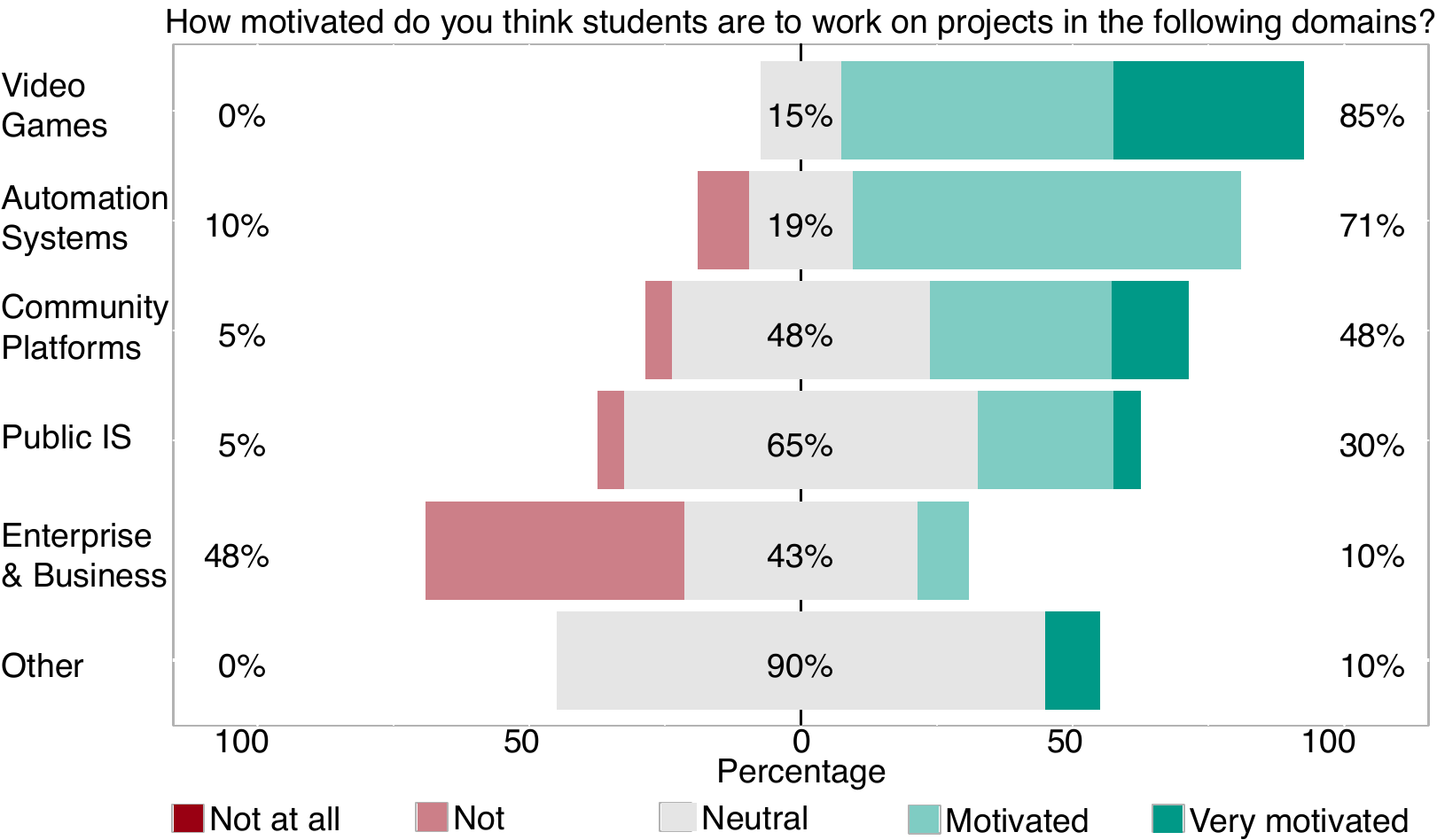}
\caption{Educators' motivation ratings by domain (RQ1).}
\Description{Educators see almost all domains as either neutral or motivating. Video games receive 85\% motivating votes, followed by automation systems (71\%), community platforms (48\%), public information systems (30\%), enterprise and business (10\%), and others (10\%). Most domains have no non-motivating votes (Video games and others) or very few (Automation systems, community platforms and public information systems). Only Enterprise and business systems receive 48\% non-motivating votes. Other system domains are seen as largely neutral (90\%).}
\label{fig:motivationEducators}
\end{figure}

We present quantitative findings on domain preferences for both students and educators, followed by qualitative themes from each group, and conclude with a comparison of their perspectives.

\subsubsection{Students Perspective} 
Figure~\ref{fig:motivationStudents} shows how motivating students find different problem domains (\emph{MO01}). 
Interestingly, none of the domains leans clearly in either direction, i.e., students have quite different opinions on what motivates them.

By comparing how motivating students of different subgroups find example domains, we see a statistically significant difference in how men and non-men view public information systems (Fisher, $p \approx 0.043$) and automation systems (Fisher, $p \approx 0.028$).
Between minorities and non-minorities, there is a statistically significant difference for video games (Fisher, $p \approx 0.038$).

To better understand \emph{why} student participants find certain domains motivating, we asked them to explain their responses in a free text field, which we analysed qualitatively. 

The analysis of the students' answers for motivation (\emph{MO02}) resulted in three major themes (shown \emph{emphasised} in the following).

\emph{Personal Interests as Primary Drivers.} 
Most students in our sample (49) explained that their motivation to work on projects depends on their own interests. They said their own interests are the main factor in deciding whether to work on a project that is related to certain domains. The content analysis revealed that the most mentioned topic here was \emph{video games}. However, while most of the participants explained having specific interests in that domain, some explicitly reported not being interested.

\emph{Relevance of Subjects.}
Twelve students explained that their motivation to work on a subject is mainly driven by what they think is relevant for society, directly, or creates value to benefit society. Our content analysis did not reveal any specific outstanding domains in the qualitative answers.

\emph{Working in Social Environments.}
One student shared that they are interested in working within social settings. This shifts the emphasis away from the subject matter itself and toward the context in which the work takes place, highlighting a preference for socially engaging environments. It indicates a desire for interaction and collaboration with others throughout the projects.

\subsubsection{Educators Perspective.}
Figure~\ref{fig:motivationEducators} shows which domains educators view as motivating for students.
Most educators in our sample believe video games and automation systems are motivating for students. Students' answers are more polarised.
Similarly, no educator in our sample believes that video games are demotivating, and only one believes that community platforms and public information systems are demotivating for students.
Two educators believe automation systems are demotivating.
Interestingly, 10 of 22 educators believe business and enterprise systems are demotivating.

According to the qualitative analysis of the educators side (\emph{MO04}), we found four themes.

\emph{Specific Students' Interests.}
Seven educators answered by explaining that students seem to be motivated by their interests and named specific domains that are the main interests of general students in their opinion/experience. Our content analysis showed that video game related words were more frequently present than others, even though there were no major differences. 

\emph{Domain-based Students' Interests.}
Five educators also agreed that the students interests are the main driver but argued that the interests depend on the main domain of the students interests. Instead of naming concrete domains, those educators generally related aspects such as the study field or the general domains of interests to the motivation of students. 

\emph{Real Subjects.}
One educator argued that students are interested in subjects that are real and complex. This suggests a need for real life examples that are not artificially created only for teaching purposes. The respective educator also explained to teach engineering students and argued that this might be a potential reason for this kind of real world focused motivation.

\emph{Own Knowledge.}
Another educator explained to have answered based on their own knowledge. Since they are not experienced with certain domains themselves they do not teach those.

\subsubsection{Comparison of Students and Educators}
We observe a statistical significant difference for students' and educators' views on public information systems (Fisher, $p \approx 0.018$), automation systems (Fisher, $p \approx 0.008$), and enterprise and business systems (Fisher, $p \approx 0.004$).
Qualitatively, both groups agree that personal interest drives motivation, though educators in our sample tend to generalise about which domains interest students, while students emphasise diverse preferences. Educators also highlighted real-world relevance less often than might be expected, possibly because they are aware of the complexity such examples can introduce~\cite{chakraborty2023we,paige2014bad}.

\summary{RQ1}{Students' interests is seen as an universal driver of motivation. Educators had strong views about which domains are motivating, whereas students' opinions vary widely.}

\subsection{RQ2: Teaching Methods and Designs}
We investigated how preferences for \emph{teaching methods} (\emph{PR02-PR06}) and \emph{assignment design choices} (\emph{DA01, DA02, DA04}) affect motivation, from the perspective of students and educators.

\subsubsection{Students Perspective}

\paragraph{Gamification (PR02-03)} 
We asked students about the role of gamification as a motivator in modelling education.

A clear majority of students stated that gamification increases their motivation (\emph{PR02}, $n=40$ strongly agree, $n=30$ agreed), with only 8 students stating that it does not affect their motivation and 4 that it decreases their motivation.
There was no statistically significant difference in this question between men and non-men (Fisher, $p \approx 0.599$) as well as between students belonging to at least one minority and students belonging to none (Fisher, $p \approx 0.568$).

When asked to explain their views, students gave several reasons for and against gamification. 
Most students in our sample (37) value how gamification may \emph{help with motivation}, and four refer to potential benefits in \emph{increasing learning outcomes}. Some students (30) report just \emph{liking gamification}. However, gamification can backfire as it can \emph{lead to frustration} (5), four highlight that some students \emph{dislike competition}, and three are concerned that gamification may \emph{distract students from learning outcomes}. Three students express how gamification \emph{feels condescending} to them.

\paragraph{Collaboration (PR04-PR05)} 
We asked students about collaborative work in modelling assignments.

The picture for collaboration practices (\emph{PR04}) is fairly balanced: 32 students stated that they prefer group projects, 29 preferred individual projects, and 21 students had no preference.
There was no statistically signifiant gender difference in this questions.

When asked to explain their views, students gave several reasons for and against collaboration. 
The \emph{balance between students who prefer to work in teams (32) and those who prefer to work solo} (29), reflects on the weight those students attribute to the \emph{benefits of discussing} (11) or \emph{cross-evaluating} their modelling attempts (4), as well as the \emph{development of teamwork skills} (3), versus concerns related to \emph{distrusting their peers} abilities and commitment to those projects (15), and \emph{fears of uneven workload balance} (4).

\subsubsection{Educators Perspective}
We also asked educators about gamification and collaboration (\emph{PR02-05}), as well as other strategies they use to encourage engagement (\emph{PR06) }and how they design assignments (\emph{DA01, DA02, DA04}).

\paragraph{Gamification (PR02-03)} 
Most educators in our sample see potential in \emph{using gamification to motivate students}, but tend to take it with a grain of salt. Some educators (4) expressed concerns about \emph{over-reliance on gamification}, whereas others (5) have no experience with it.
A few educators (3) believe students like gamification, but others (2) highlight that it does not work for all students: 
\interviewquote{I think gamification helps to motivate students, as long as there is not a strong competitive factor (e.g. public leaderboards). If there is, students have widely different reactions, thus affecting motivation differently depending on the student.}{~\textsuperscript{ED101}} 

\paragraph{Collaboration (PR04-PR05)} 
Most educators reported that they \emph{believe that students prefer to work in groups} (17). They identify \emph{pedagogical benefits} to having students work in teams, as students get to \emph{discuss their opinions} on the models they create (14) and \emph{learn more} in the process. These discussions highlight that there is often more than one possible modelling solution to a given problem (6), which supports the argument for group work. \emph{Scalability}, both in terms of modelling challenges and evaluation work by educators, favours collaborative teams. Educators recognise the benefits of including individual modelling assignments as a way of \emph{fostering autonomy} (2) and \emph{preventing uneven workload among students} (1).

\paragraph{Encouraging students to model (PR06)} 
We asked educators about encouraging students to engage with modelling assignments.

Some educators (5) make these assignments mandatory as \emph{part of the evaluation process}. One educator suggested that a further motivation could be that the written exam topic could be the assigned project.
Three educators highlighted the relevance of using \emph{interesting domains}, e.g., by allowing students to suggest the application domains they want to work on.
A complementary approach is to highlight the modelling's \emph{relevance to industry}, e.g., by drawing on their own experience.
Two educators prefer highlighting the different \emph{modelling alternatives}, while two others stress the benefits of providing \emph{personalised feedback} to students.\\

In addition, we asked educators how they select domains (\emph{DA01}), which domains they choose (\emph{DA02}), and how students' interests influence this choice (\emph{DA04}). These questions focused on educators as they are responsible for designing assignments.

\paragraph{Choosing modelling domains (DA01)}
Educators mention students' \emph{interest} as the most frequently cited criterion for selecting a domain for a modelling assignment (8). The domain's \emph{alignment with the student profile} is also considered important (5). One educator uses both as they mentioned to \emph{``choose a domain that fits within their degree [...] and their interest to engage them as much as possible.''}~\textsuperscript{ED86}
General \emph{relevance and suitability} are mentioned by six educators each. \emph{Fitness to the learning goals} is mentioned by 4.

\paragraph{Used domains (DA02)}
There is a wide variety of specific domains in use, with banking (4) and university (4) being the most frequent. More broadly, educators mentioned domains related to travel and transports (11), organisational and business processes (10), education and knowledge (9), digital media and entertainment (9), cyber-physical systems (6), and e-commerce and shopping (2).

\paragraph{Student interest and motivation influence on modelling assignments design (DA04).}
The relevance educators assign to students' interests and motivation is mixed. Some educators consider it important (7), or relatively important (3). Others find it not important (4). When considering students' interests, some educators mention domain familiarity (3) and fitness to learning goals (3) as a criterion for their choice, followed by using a tangible domain (2).

\subsubsection{Comparison of Students and Educators}
Students believe that gamification increases their motivation more than educators do (Fisher, $p < 0.001$). Qualitatively, students are more enthusiastic about gamification overall, though both groups recognise it can backfire, particularly when it becomes overly competitive.

Regarding collaboration, there is a notable mismatch: most educators in our sample assume students prefer group work, yet students are evenly split in their preferences. Students who favour individual work often cite concerns about peers' commitment and uneven workload, while educators emphasise the pedagogical benefits of discussion and learning from multiple perspectives.

\summary{RQ2}{Gamification may increase motivation, but could backfire if poorly designed. Educators view collaboration more positively and select relatable domains to engage students.}

\subsection{RQ3: Inclusion and Bias}
We investigated how students perceive inclusiveness in assignment domains (\emph{IN02–IN04}) and how educators consider diversity and bias in domain selection (\emph{EI01–EI03}).

\subsubsection{Students Perspective}

\paragraph{Experiences of Exclusion (IN02)}  
Five main themes emerged from students’ experiences: no exclusion, exclusion occurred due to unfamiliar domains, representational issues (e.g., gender or queerness), assignment settings, and accessibility barriers. 

Most students (39) reported no exclusionary experiences, e.g., due to the assignments having no connection to real life. 

However, several students described exclusion linked to domain familiarity, e.g. if they had to build \emph{``a project which included game elements and if you don't game then you were at a disadvantage.''}~\textsuperscript{ST5}
This unfamiliarity can stem from culturally specific references and backgrounds when the assignment is 
\interviewquote{based on a game that's well-known mainly in Europe. I even read through the entire Wikipedia article about the game rules, but I still couldn't fully understand it}{\textsuperscript{ST75}}

Other students (9) pointed to representational aspects: 
\interviewquote{When thinking about the LGBTQ+ community... I can recall some assignments only referring to male and female, which could be a bit disregarding of other genders}{~\textsuperscript{ST19}}

Further, students (3) highlighted not the topic but the assignment setting itself as exclusionary, e.g, if you prefer to work alone:  \interviewquote{A big group where work is conducted in class is pretty exclusionary to people with very big motor disabilities}{~\textsuperscript{ST32}}

\paragraph{Impact of Domain Choice on Inclusiveness (IN03)}  
Students highlighted four key themes: no impact (18), ambivalent opinions (12), and domain choice affects motivation depending on their background and problematic content (24). 

One of the reasons why students think of the impact on inclusiveness is the lack of sense of belonging since if all
\interviewquote{assignments were to be specific to one type of person, others could feel demotivated and think this wasn't the course for them}{~\textsuperscript{ST28}} 

This is also due to the diversity in backgrounds as \emph{``a topic that is familiar to some may be difficult for others.''}~\textsuperscript{ST54}
Concerns about biased or problematic domains were also raised: 
\interviewquote{In some domains there are inhered misogyny or phobia, in a modelling assignment it could be hard to get rid of them}{~\textsuperscript{ST51}}

\paragraph{Impact on Learning Success (IN04)}  
Students described three themes regarding learning outcomes (\emph{IN04}): familiarity or interest in a domain improves understanding, unfamiliar domains can reduce engagement, and motivation mediates the effect. 
Most students (33) reported positive impact due to interesting domains: 
\interviewquote{Familiar or interesting domains boost understanding, abstract or unfamiliar ones can hinder it. Helpful: Healthcare, transport, education. Harder: Physics simulations, finance models}{~\textsuperscript{ST57}}

Some domains increased motivation (e.g., Minecraft, sports), but individual preferences varied. Most students emphasised real-world relevance since if the domain is close to their life, \emph{``it could be interesting to look at familiar things from a new perspective''}.{~\textsuperscript{ST51}}

Ten students felt there was little to no impact at all if the assignment is well-designed and explained since they 
\emph{``try to understand the underlying concepts than the specific application''.}~\textsuperscript{ST42}

\paragraph{Preferred Domains for Diverse Groups (IN01).}   
Three themes emerged regarding preferred domains (\emph{IN01}): students favour societal relevance, everyday experiences that all students can relate to, and the ability to choose or personalise domains. 

Students suggested broadly accessible and socially relevant domains, including societal topics (culture, mental health, climate; 19), community platforms (12), education (9), everyday student life (5), games which are not related to \emph{gamer culture} (3).

Since there is often the risk of \emph{othering} students who belong to an under-represented group, students emphasised choice: 
\interviewquote{I think that's really hard as we are all different. I believe that the approach ‘choose your own topic/theme’ is the best one.}{~\textsuperscript{ST28}}  

\subsubsection{Educators Perspective.} 

\paragraph{Considering Diversity in Domain Selection (EI01)} 
Educators’ practices (\emph{EI01}) revealed three main patterns: some explicitly consider diversity to increase accessibility and challenge stereotypes (7), others implicitly account for diversity via familiar contexts or student autonomy (4), and some do not consider diversity at all (5). 

When explicitly considering diversity, strategies included using familiar domains for international students, avoiding conflict-prone domains, and selecting domains that challenge binary thinking.

When addressing implicit diversity, educators draw on everyday contexts relatable to all, use study-related domains, or allow students to choose.  
As one third do not consider diversity, inclusive assignment design is unevenly integrated into teaching practice.

\paragraph{Observed Differences in Student Preferences (EI02)} 
Three themes emerged (\emph{EI02}): educators who notice differences adjust for personal interests or background (3), some acknowledge potential differences but do not act on them (2), and most are unaware of differences or bias (11). 
For example, one observed gender-specific patterns: 
\interviewquote{Some games appeal more to a male audience, e.g., a spaceship strategy game. Yet, most games appeal to all audiences. [...] To be more inclusive towards women in computer science education, a colleague argued that projects for sharing, ecological or sociological improvements are more appealing to women}{~\textsuperscript{ED152}}

\paragraph{Suggestions for Inclusive Modelling Education (EI03)}
Four themes guided educators’ suggestions (\emph{EI03}): providing a variety of domains and task types (6), fostering interaction with diverse perspectives (2), using accessible tools, and considering student backgrounds. 

The most common recommendation is the choice of task and \emph{variety of domains}, yet, apply it with caution:
\interviewquote{But for certain physical disabilities (visual impairment, physical impairment of hands) modelling is quite hard because of the tooling not adapted to these students.}{~\textsuperscript{ED70}} 
Another suggested a gradual sequence of assignments: 
\interviewquote{We need small, comprehensive examples from the student's life or experience, an inclusive medium example to train, and finally a real-world example. Only the small and medium case must be inclusive, whereas the big one should prepare the students for their job}{~\textsuperscript{ED152}}  

Two educators emphasised interaction and engagement with diverse perspectives, suggesting, for instance: \interviewquote{Having students interact with people (or at least personas) of disadvantage categories as 'clients' of their projects}{~\textsuperscript{ED95}}

Considering students' background (e.g., cultural, educational, socio-economic), educators warned against unfamiliar domains:
\interviewquote{Try not to pick domains that you yourself like, but students may not know/like, or which may even discriminate. Classical example: Booking a hotel – some students may never have stayed in a hotel since they cannot afford it}{~\textsuperscript{ED143}}

Two educators were sceptical about addressing inclusivity at course level, seeing it as a broader institutional responsibility.

\subsubsection{Comparison of Students and Educators}
Students and educators in our sample agree that interest influences motivation and inclusiveness. However, students prioritise choice and societal relevance, while educators focus more on familiarity and manageability of domains. Students more clearly identify exclusion risks (e.g., unfamiliar references, economic assumptions, limited representation), whereas many educators are unaware of such issues or do not actively address them, suggesting a misalignment between intentions and students’ lived experiences.

\summary{RQ3}{Most students in our sample find assignment domains mainly inclusive. Students value choice and societal relevance, while educators’ awareness and strategies vary.}

\subsection{RQ4: Feedback}
We investigated how students perceive feedback (\emph{FE03}), how educators provide it (\emph{FE01}), and how they evaluate their course quality (\emph{FE02}) to explore the dynamics of feedback in modelling education. 

\subsubsection{Students Perspective} 
Students' responses (\emph{FE03}) clustered around three major themes: perceived positive impact, conditional impact, and lack of visible impact. Interestingly, they all interpreted feedback as course quality feedback, not on their learning.

\emph{Positive Impact.}
Most students (32) believed feedback enhances course quality, emphasising  motivational and practical benefits: 
\interviewquote{Without any feedback, the teaching person will think that everything is perfect and not change anything. Feedback might motivate to change something and enhance the quality of the course.}{~\textsuperscript{ST41}}  
The positive impact allows also for hearing diverse views since \emph{“many opinions might give new perspectives”}.~\textsuperscript{ST56}

\emph{Conditional Impact.}
Several students (19) expressed more nuanced views by linking the effectiveness to timing, lecturer responsiveness, and the visibility of integration and peer agreement: 
\interviewquote{it can also become frustrating when 1. feedback is not taken at a point when it can still improve the students taking the course at that moment, and 2. when we can see that feedback from previous years was clearly not taken into consideration, as some issues remain recurring.}{\textsuperscript{ST35}}

Some students felt their input mattered only if peers agreed or if surveys were representative. In addition, some stressed that impact depends on personal fit: 
\interviewquote{My feedback would improve the quality for me. If none share my opinion it'd probably decrease.}{~\textsuperscript{ST28}}

\emph{Lack of Impact.}
Several students (13) doubted that feedback influenced teaching, pointing to entrenched course structures or low institutional priority of teaching: \interviewquote{Teaching is just a side task for the teachers and professors. Normally they just stick to the old course structures and slides.}{~\textsuperscript{ST51}}

\subsubsection{Educators Perspective.} 
Educators reported three main approaches for their practices (\emph{FE01, FE02}).

\emph{Formative and personalised feedback.} Educators (6) emphasised one-on-one discussions and or small-group feedback, often provided in consultation hours or practical sessions: \interviewquote{I periodically hold consultation sessions where students can discuss their group assignments. Practical sessions are also used to provide feedback as needed.}{~\textsuperscript{ED07}} 

\emph{Scalable mechanisms.} Several educators (5) highlighted iterative cycles, peer review, or automated platforms, particularly for larger cohorts. One approach combined automation and peer processes: 
\interviewquote{Small single-person assignments receive automatic and individual feedback via an exercise platform. For the group project, we employ a double-blind peer-review process, where each group evaluates the models of two other groups.}{~\textsuperscript{ED18}} 

\emph{Course evaluation.} Educators (12) typically relied on standardised student surveys and university-wide processes to assess quality. Some highlighted long-term course refinement.

\subsubsection{Comparison of Students and Educators}
Both groups agree that feedback is important, but the perception of its effectiveness differs. While educators actively provide feedback through multiple channels, students sometimes perceive it as ineffective if not visibly acted upon. This gap between provision and perceived impact highlights a potential risk: students may feel their voices do not matter, especially those from under-represented groups. 

\summary{RQ4}{Students value feedback mainly when visibly acted upon. Educators provide it regularly, but perception gaps can reduce its motivational impact.}

\section{Discussion}
\label{sec:discusiion}
\begin{table}[t]
\caption{Recommendations for software modelling education.}
\label{tab:recommendations}
\centering
\small
\begin{tabular}{>{\raggedright\arraybackslash}p{2.4cm}>{\raggedright\arraybackslash}p{5.6cm}}
\toprule
Goal & Recommendation \\ 
\midrule
Use students' interests & Offer opportunities to choose the project domains by students, if possible (\emph{RQ1, RQ2}). \\ 
Show societal relevance & Link domains to real-world, socially meaningful problems, not just technical examples (\emph{RQ1, RQ3)}. \\
Use gamification wisely & Keep it engaging but avoid overemphasis on competition (\emph{RQ2)}.\\
Support collaboration & Scaffold teamwork, discuss \emph{fair} work distribution, and use mechanisms like blind peer evaluation (\emph{RQ2}).\\
Avoid narrow examples &  Do not rely on culturally specific or stereotypical domains; prefer everyday contexts (\emph{RQ1, RQ3}).\\
Make feedback visible & Collect feedback iteratively and show students how their input influences teaching, giving minorities a voice (\emph{RQ4}). \\
\bottomrule
\end{tabular}
\end{table}

Our study provides insights into students’ motivations in software modelling courses, highlighting how domain diversity, the inclusiveness of assignments, and the feedback they receive jointly influence student engagement.

\subsection{Synthesis of Results}
Our findings suggest a mismatch between educators' and students' perceptions of what makes modelling motivating. Educators hold more uniform and narrow views on suitable domains (RQ1) and teaching methods such as gamification or collaboration (RQ2). 
In contrast, students report a wider variety of interests and are cautious about approaches that feel forced or overly competitive. What consistently drives motivation is when assignments connect to students’ personal interests, or to socially meaningful, everyday domains, which resonates with prior work on autonomy and relevance in learning \cite{trenshaw2016using,cevikbas2023advantages,Weber2024}. 
This reframes inclusiveness beyond technical access: while most students report no overt exclusion (RQ3), those who do point to culturally specific references and economic assumptions in domain choices educators do not anticipate, connecting inclusivity concerns (RQ3) to domain and teaching method selection practices explored in RQ1 and RQ2. Students feel more included when domains reflect their lived realities, echoing calls for diversity-rich curricula \cite{ramdas2025creating,burgstahler2009universal}. 
The feedback gaps identified in RQ4 may partially explain the persistent educator-student misalignments observed across RQ1-3: while educators design structured mechanisms, students value it only when they can see tangible effects on their learning, supporting earlier observations on the importance of formative, responsive practices \cite{chakraborty2023we,kienzle2024requirements}. 

Inclusive modelling education depends on creating assignments that make students feel represented, not just technically competent, in line with findings from Computer Science education~\cite{Michaelis2022TCE,Briesen2025SIGSETS}. It requires educators to view inclusiveness as more than access, instead designing tasks that acknowledge students’ voices, backgrounds, and aspirations. Overall, our findings highlight that effective modelling education must balance educator intentions with student experiences, leveraging domain diversity, inclusive practices, and timely feedback to increase both motivation and engagement.

\subsection{Connecting Educators and Students}
\Cref{tab:recommendations} summarises our practical recommendations for more inclusive software modelling education.

While demographic factors are often reported, the diversity of domains students engage with remains under-explored~\cite{dutta2023diversity}. Our results show that engagement is strongly shaped by individual domain preferences, with many students motivated by socially relevant problems that educators often overlook (\emph{RQ1}, \emph{RQ3}). This suggests that educators should offer choice and link tasks to real-world issues rather than relying on assumedly motivating technical topics, echoing prior work on ``diversity in problem-solving''~\cite{hyrynsalmi2025challenges}.

Gamification is another challenge: most students in our sample enjoy it, but some find competitive elements discouraging (\emph{RQ2}), in line with earlier findings~\cite{hanus2015assessing}. Educators should use a broader set of game elements to enhance engagement, ideally integrated with collaborative work. Collaboration itself divides opinions: while educators highlight its benefits, some students distrust peers or fear uneven workload distribution (\emph{RQ2}), a pattern that resonates with earlier studies~\cite{iacob2019exploring}. This points to a need for better communication and scaffolding of collaborative tasks.

Students value course feedback when they see that educators act upon it (\emph{RQ4}). End-of-semester surveys rarely demonstrate impact and may silence minority perspectives. Iterative feedback with visible follow-up during courses can address this gap. 
By offering autonomy in domain choice, applying gamification thoughtfully, scaffolding collaboration, and ensuring responsive feedback, educators can foster more inclusive and motivating modelling education.

\section{Conclusions and Future Work}
\label{sec:conclusion}
Our study shows that motivation, domain choice, inclusivity, and feedback are  interconnected in software modelling courses. From a students perspective, socially relevant, relatable, and self-selected domains as well as well-executed gamification enhance engagement, while visible and responsive feedback is essential for inclusive participation. Educators’ assumptions about student interests and reliance on standardised surveys can limit motivation and inclusion.

This is the first empirical study comparing educators' and students' perspectives on these factors in modelling education.
Our findings are based on self-reported perspectives, which are important indicators related to motivation. However, actual learning might be affected by further factors. 
Intervention studies should therefore test whether incorporating our recommendations \emph{actually} improves student engagement and outcomes, particularly for underrepresented groups. Finally, cross-cultural replication helps determine whether our findings from predominantly European contexts generalise to other educational settings, or whether culturally-specific approaches to inclusive modelling education are needed.

\section*{Acknowledgments}
This study was developed as a part of the GI-Dagstuhl Seminar 23473: Human Factors in Model-Driven Engineering. We thank Schloss Dagstuhl for the opportunity to conduct the seminar. The work is funded by the Deutsche Forschungsgemeinschaft (DFG, German Research Foundation) – Project-ID 414984028 – SFB 1404 FONDA, and NOVA LINCS (\href{https://doi.org/10.54499/UID/04516/2025}{\url{UID/04516/2025}}) with the financial support of FCT.IP.

\newpage
\section*{Data Availability}
We provide the anonymised survey data, including questionnaires and introductory text, in our replication package~\cite{companionSite}.

\balance
\bibliographystyle{ACM-Reference-Format}
\bibliography{references}

\end{document}